\documentclass[preprint,showpacs,showkeys,preprintnumbers,amsmath,amssymb,superscriptaddress]{revtex4}

\usepackage{natbib}
\usepackage{graphicx}

\begin{document}

\title{Entropic functionals of Laguerre polynomials and complexity properties of the half-line Coulomb potential}

\author{P. S\'anchez-Moreno}
\email{pablos@ugr.es}
\affiliation{Instituto Carlos I de F\'{\i}sica Te\'{o}rica y Computacional, University of Granada, Granada, Spain}
\affiliation{Departamento de Matem\'{a}tica Aplicada, University of Granada, Granada, Spain}

\author{J.J. Omiste}
\email{omiste@ugr.es}

\author{J.S. Dehesa}
\email{dehesa@ugr.es}
\affiliation{Instituto Carlos I de F\'{\i}sica Te\'{o}rica y Computacional, University of Granada, Granada, Spain}
\affiliation{Departamento de F\'{\i}sica At\'{o}mica, Molecular y Nuclear, University of Granada, Granada, Spain}

\date{\today}

\begin{abstract}
The stationary states of the half-line Coulomb potential are described by quantum-mechanical wavefunctions which are controlled by the Laguerre polynomials $L_n^{(1)}(x$).
Here we first calculate the $q$th-order frequency or entropic moments of this quantum system, which is controlled by some entropic functionals of the Laguerre polynomials. These functionals are shown to be equal to a Lauricella function $F_A^{(2q+1)}\left(\frac{1}{q},\ldots,\frac{1}{q},1\right)$ by use of the Srivastava-Niukkanen linearization relation of Laguerre polynomials. The resulting general expressions are applied to obtain the following information-theoretic quantities of the half-line Coulomb potential: disequilibrium, Renyi and Tsallis entropies.
An alternative and simpler expression for the linear entropy is also found by means of a different method. Then, the Shannon entropy and the LMC or shape complexity of the lowest and highest (Rydberg) energetic states are explicitly given; moreover, sharp information-theoretic-based upper bounds to these quantities are found for general physical states. These quantities are numerically discussed for the ground and various excited states. Finally, the uncertainty measures of the half-line Coulomb potential given by the information-theoretic lengths are discussed.
\end{abstract}

\pacs{02.30.Gp; 03.65.-w; 89.70.Cf}

\keywords{Half-line Coulomb potential; information theory; linearization of Laguerre polynomials; entropic moments; Renyi entropy; Tsallis entropy; Shannon entropy; LMC complexity; Rydberg states; information-theoretic lengths}

\maketitle

\section{Introduction}

The application of information-theoretic ideas and techniques to the study of multielectronic systems in the last few years has provided to be an interesting new channel of cross-fertilization between atomic and molecular physics and the physics of information. This line of inquiry has recently led to numerous researchers to calculate the measures of information for various quantum-mechanical potentials of a specific analytic form \cite{patil:jpb07,sen:jcp05,patil:pla07,patil:ijqc07,patil:pla07_2,dehesa:jmp98,atre:pra04,dehesa:mp06,nagy:pla09} as well as for hydrogenic systems with standard and non-standard dimensionalities \cite{patil:jpb07,lopez:09,dehesa:pra94,dehesa:jmp98,dehesa:ijqc09,lopezrosa:pa09}, circular membranes \cite{dehesa:ijbc02}, confined systems \cite{omiste:jmp09,sen:aqc09} and many-electron systems \cite{guevara:jcp03,guevara:pra03,plastino:pa06,plastino:ptps06,angulo:pra94}.

In this work we determine some single (entropic moments, Renyi and Tsallis entropies) and composite (LMC or shape complexity) information-theoretic-measures of the stationary states of the one-dimensional (1D) hydrogenic atom, which are described by the physical solutions $(E,\psi)$ of the Schr\"odinger equation of an electron moving in the half-line Coulomb potential, $V(x)=-Ze^2/x$ for $x>0$ and infinity for $x<0$. The symbols $e$ and $Z$ denote the electronic and the nuclear charges, respectively. This system has been used in the modelling of the dynamics of surface-state electrons in liquid helium \cite{platzman:s99,nieto:pra00,dykman:prb03} and the behaviour of atoms irradiated by external fields \cite{leopold:prl78,richards:jpb87,leopold:jpb88}. Its usefulness has been shown to analyze the main features of revival and functional revival phenomena \cite{veilande:jpb07}, to describe some aspects of Rydberg atoms \cite{mayle:prl07} and to be possible candidates for a two-qubit quantum gate \cite{jaksh:prl00} in quantum computation. Moreover, some authors claim to have produced it at the laboratory \cite{stokey:pra03}.

The energetic eigenvalues $E_n$ and the corresponding eigenfunctions $\psi_n(x)$ of the 1D hydrogenic atom are known \cite{nieto:pra00} to have the following form:
\[
E_n=-\frac{Z^2}{n^2},\quad \psi_n(x)=\sqrt{\frac{Z}{n^3}}  \sqrt{t\omega_1(t)} L_{n-1}^{(1)}(t),\; n=1,2,\ldots 
\]
where we have used atomic units (Planck's constant = electron mass = electron charge = 1) and the symbol $t=2Zx/n$, and $L_k^{(\alpha)}(t)$ denotes the Laguerre polynomials \cite{temme_96}, which are orthogonal with respect to the weight function $\omega_\alpha(t)=t^\alpha e^{-t}$ on the interval $[0,\infty)$; that is,
\[
\int_0^\infty L_n^{(\alpha)}(t)L_m^{(\alpha)}(t)\omega_\alpha(t)dt=\frac{\Gamma(n+\alpha+1)}{n!}\delta_{n,m}.
\]
Then, the quantum-mechanical probability density of the system is given by
\begin{equation}
\rho_n(x)=|\psi_n(x)|^2=\frac{Z}{n^3}t\omega_1(t) \left[L_{n-1}^{(1)}(t)\right]^2.
\label{eq:rho}
\end{equation}

Recently, the spreading and uncertainty-like properties of the 1D hydrogenic atom have been studied \cite{omiste:jmp09} by means of the Heisenberg measure (i.e. the root-mean-square or standard deviation) and far beyond, by use of the logarithmic measure and the information-theoretic lengths of Renyi, Shannon and Fisher types \cite{hall:pra99}. However, the calculation of the Renyi lengths could not then be performed because the computation of the frequency or entropic moments of the system, defined as
\begin{eqnarray}
W_q[\rho_n]&:=&\int_0^\infty \left[\rho_n(x)\right]^q dx \equiv \left\langle\left[\rho_n(x)\right]^{q-1}\right\rangle
\label{eq:wq_a}\\
&=&\left(\frac{Z}{n^3}\right)^q \frac{n}{2Z}\mathcal{I}_q\left[L_{n-1}^{(1)}\right],\; q=0,1,\ldots,
\label{eq:wq_b}
\end{eqnarray}
have not yet been accomplished in an analytical way. The reason for this situation is that the entropic functionals
\begin{equation}
\mathcal{I}_q\left[L_{n}^{(1)}\right]:=\int_0^\infty t^q \left\{\omega_1(t)\left[L_{n}^{(1)}(t)\right]^2\right\}^qdt
\label{eq:Iq}
\end{equation}
have not yet been found, except for the ground state ($n=1$).

In this work we will show that these integrals of Laguerre polynomials can be expressed in terms of the quantum number $n$ and the integer order $q$ via the Lauricella function $F_A^{(2q+1)}(\frac{1}{q},\ldots,\frac{1}{q},1)$ \cite{srivastava_85}. This is done in Section \ref{sec2}. This general mathematical result is basic to compute the entropic moments $W_q[\rho_n]$ of any hydrogenic stationary state, as given by Eq. (\ref{eq:wq_b}), and various physical and information-theoretic quantities closely related to them such as the Renyi and Tsallis entropies (see Section \ref{sec3}). In Section \ref{sec4}, we use an alternative method which allows us to find a simpler explicit expression for the disequilibrium and linear entropy of a generic quantum state of our system. Then, the Shannon entropy and the shape complexity of the states at the two energetic extremes of the spectrum are explicitly given in Section \ref{sec5}. Moreover, in Section \ref{sec6}, sharp upper bounds to these two quantities are obtained for general physical states.
Finally, the uncertainty measures of the half-line Coulomb potential given by the  information-theoretic spreading lengths \cite{hall:pra99} are discussed in Section \ref{sec7} and some conclusions are given.

\section{Entropic functionals of Laguerre polynomials }
\label{sec2}

Here we calculate the entropic functionals $\mathcal{I}_q\left[L_{n-1}^{(1)}\right]$ of integer order $q$ of the Laguerre polynomial $L_{n-1}^{(1)}(t)$ defined by Eq. (\ref{eq:Iq}); that is
\begin{equation}
\mathcal{I}_q\left[L_{n-1}^{(1)}\right]=\int_0^\infty t^{2q} e^{-qt} \left[L_{n-1}^{(1)}(t)\right]^{2q}dt.
\label{eq:Iq2}
\end{equation}
To do that we use the following linearization relation of products of Laguerre polynomials \cite{srivastava:mcm03}
\[
x^\mu L_{m_1}^{(\alpha_1)}(x_1z)\cdots L_{m_r}^{(\alpha_r)}(x_rz)=\sum_{k=0}^\infty \theta_k(\mu;x_1,\ldots,x_r)L_k^{(\alpha)}(z),
\]
where the coefficients $\theta_k(\mu;x_1,\ldots,x_r)$ can be expressed as
\begin{multline*}
\theta_k(\mu;x_1,\ldots,x_r)=(\alpha+1)_\mu \binom{m_1+\alpha_1}{m_1}\cdots\binom{m_r+\alpha_r}{m_r}\\
\times F_A^{(r+1)}\left(\alpha+\mu+1,-m_1,\ldots,-m_r,-k;\alpha_1+1,\ldots,\alpha_r+1,\alpha+1;x_1,\ldots,x_r,1\right)
\end{multline*}
in terms of Lauricella's hypergeometric functions of $r+1$ variables \cite{srivastava_85}. The Pochhammer symbol $(a)_\alpha=\frac{\Gamma(a+\alpha)}{\Gamma(a)}$ and the binomial number $\binom{a}{b}=\frac{\Gamma(a+1)}{\Gamma(b+1)\Gamma(a-b+1)}$. For the special case $\mu=2q$, $r=2q$, $\alpha_1=\ldots=\alpha_r=1$, $m_1=\ldots=m_r=n-1$, $x_1=\ldots=x_r=1/q$, $x=z=qt$, and $\alpha=0$, this general relation immediately yields the following result
\begin{equation}
(qt)^{2q}\left(L_{n-1}^{(1)}(t)\right)^{2q}=\sum_{k=0}^{2qn}\theta_k\bigg(2q;\underbrace{\frac{1}{q},\ldots,\frac{1}{q}}_{2q}\bigg)L_k^{(0)}(qt),
\label{eq:qtL}
\end{equation}
where the linearization coefficients are given by
\[
\theta_k\bigg(2q;\underbrace{\frac{1}{q},\ldots,\frac{1}{q}}_{2q}\bigg)=(2q)! n^{2q}
F_A^{(2q+1)}\bigg(2q+1,\underbrace{-n+1,\ldots,-n+1}_{2q},-k; \underbrace{2,\ldots,2}_{2q},1;\underbrace{\frac{1}{q},\ldots,\frac{1}{q}}_{2q},1\bigg).
\]
From Eqs. (\ref{eq:Iq2}) and (\ref{eq:qtL}), one has that
\begin{multline*}
\mathcal{I}_q\left[L_{n-1}^{(1)}\right]=\frac{1}{q^{2q}} \int_0^\infty (qt)^{2q}e^{-qt} \left(L_{n-1}^{(1)}(t)\right)^{2q}dt\\
= \frac{1}{q^{2q}}\int_0^\infty e^{-qt} \sum_{k=0}^{2qn}\theta_k\left(2q;\frac{1}{q},\ldots,\frac{1}{q}\right)L_k^{(0)}(qt)dt.
\end{multline*}

Taking into account the orthogonality relation of the Laguerre polynomials, only the term with $k=0$ gives a non-vanishing contribution to this summation, so that
\begin{equation}
\mathcal{I}_q\left[L_{n-1}^{(1)}\right]=\frac{1}{q^{2q}}\int_0^\infty e^{-qt} \theta_0\left(2q;\frac{1}{q},\ldots,\frac{1}{q}\right)L_0^{(0)}(qt)dt=\frac{1}{q^{2q+1}} \theta_0\left(2q;\frac{1}{q},\ldots,\frac{1}{q}\right),
\label{eq:Iq_th0}
\end{equation}
where the function $\theta_0$ has the following expression
\begin{multline}
\theta_0\left(2q;\frac{1}{q},\ldots,\frac{1}{q}\right)\\= (2q)! n^{2q}
F_A^{(2q+1)}\left(2q+1,-n+1,\ldots,-n+1,0;2,\ldots,2,1;\frac{1}{q},\ldots,\frac{1}{q},1\right)\\
=(2q)! n^{2q} \sum_{m_1=0}^{n-1}\ldots\sum_{m_{2q}=0}^{n-1} \frac{(2q+1)_{m_1+\ldots+m_{2q}}(-n+1)_{m_1}\cdots(-n+1)_{m_{2q}}}{(2)_{m_1}\cdots(2)_{m_{2q}}}
\frac{q^{-m_1}\cdots q^{-m_{2q}}}{m_1!\cdots m_{2q}!}.
\label{eq:th0}
\end{multline}

Then, we have that the $q$th-order entropic functional $\mathcal{I}_q[L_{n-1}^{(1)}]$ is given by
\begin{multline}
\mathcal{I}_q\left[L_{n-1}^{(1)}\right]=\frac{(2q)!n^{2q}}{q^{2q+1}}\\
\times\sum_{m_1=0}^{n-1}\ldots\sum_{m_{2q}=0}^{n-1} \frac{(2q+1)_{m_1+\ldots+m_{2q}}(-n+1)_{m_1}\cdots(-n+1)_{m_{2q}}}{(2)_{m_1}\cdots(2)_{m_{2q}}}
\frac{q^{-m_1}\cdots q^{-m_{2q}}}{m_1!\cdots m_{2q}!}.
\label{eq:Iqgeneral}
\end{multline}

Though we can perform the $m_1$-sum as
\begin{multline*}
\sum_{m_1=0}^{n-1} \frac{(2q+1)_{m_1+\cdots+m_{2q}}(-n+1)_{m_1}}{(2)_{m_1}q^{m_1}m_1!}\\
=(2q+1)_{m_2+\cdots+m_{2q}}\sum_{m_1=0}^{n-1} \frac{(2q+1+m_2+\cdots+m_{2q})_{m_1}(-n+1)_{m_1}}{(2)_{m_1}q^{m_1}m_1!}\\
=(2q+1)_{m_2+\cdots+m_{2q}} \,_2F_1
\left(\left.
\begin{array}{c}
1-n,2q+1+m_2+\cdots+m_{2q}\\
2
\end{array}
\right|\frac{1}{q}\right),
\end{multline*}
to go analytically beyond is a difficult task, partially because the $m_2$-sum
\begin{eqnarray*}
\sum_{m_2=0}^{n-1} \frac{(2q+1)_{m_2+\cdots+m_{2q}}(-n+1)_{m_2}}{(2)_{m_2}q^{m_2}m_2!} \,_2F_1
\left(\left.
\begin{array}{c}
1-n,2q+1+m_2+\cdots+m_{2q}\\
2
\end{array}
\right|\frac{1}{q}\right).
\end{eqnarray*}
is a formidable task. Nevertheless, we can use the general formula (\ref{eq:Iqgeneral}) to evaluate the entropic moments of the Laguerre polynomials $L_k^{(1)}(t)$ of lowest degrees. In particular, for $n=1$ we have that $\theta_0=1$ and then
\begin{equation}
\mathcal{I}_q\left[L_0^{(1)}\right]=\frac{\Gamma(2q+1)}{q^{2q+1}}.
\label{eqn:iq0}
\end{equation}

Moreover, for $n=2$ one has that
\begin{eqnarray}
\nonumber
\mathcal{I}_q\left[L_1^{(1)}\right]&=&\frac{(2q)!2^{2q}}{q^{2q+1}}\sum_{m_1=0}^{1}\ldots\sum_{m_{2q}=0}^{1} \frac{(2q+1)_{m_1+\cdots+m_{2q}}(-1)_{m_1}\cdots(-1)_{m_{2q}}}{(2)_{m_1}\cdots(2)_{m_{2q}}} \frac{q^{-m_1}\cdots q^{-m_{2q}}}{m_1!\cdots m_{2q}!}\\
\nonumber
&=&\frac{(2q)!2^{2q}}{q^{2q+1}} \sum_{j=0}^{2q}\binom{2q}{j}\frac{(2q+1)_j (-1)^j}{2^j q^j}\\
&=&\frac{(2q)!2^{2q}}{q^{2q+1}}\sqrt{\frac{2q}{\pi}}e^{-q} i K_{2q+\frac12}(-q)
\label{eqn:i1}
\end{eqnarray}
where $K_\nu(z)$ is the MacDonald or Basset function (also called by modified Bessel function of the third kind and modified Hankel function) \cite{spanier_09,temme_96}.

\section{Entropic moments and Renyi and Tsallis entropies of the half-line Coulomb potential}
\label{sec3}

Here we express the entropic moments of the physical states $\psi_n(x)$ of the half-line Coulomb potential in terms of its associated quantum number $n$, as well as the two most conspicuous information entropies related to them: the Renyi and Tsallis entropies.

Differently from the ordinary or power moments (e.g., moments around the origin, central moments,...), the frequency or entropic moments \cite{kendall_69} of the probability density $\rho_n(x)$, $W_q[\rho_n]$, defined by Eq. (\ref{eq:wq_a}) measure the total extent in which the probability is spread without respect to any specific point of its support interval. They are fairly efficient in the range where the ordinary moments are fairly inefficient \cite{shenton:b51,romera:jmp01}. In quantum theory the entropic moments have been used as (indirect) measures of uncertainty (as extensively investigated by Maassen and Uffink \cite{maassen:prl88}; see also Ref. \cite{angulo:jmp00}) and have been shown to be related to some fundamental quantities of the system \cite{dehesa:pra89,dehesa:pra88,liu:pra99}. Furthermore, these moments are closely connected to the following information-theoretic quantities: the Renyi entropies \cite{renyi_70}
\begin{equation}
R_q[\rho_n]:=\frac{1}{1-q}\ln W_q[\rho_n],\; q>0,\; q\neq 1,
\label{eq:renyi_n}
\end{equation}
and the Tsallis entropies \cite{tsallis:jsp88}
\begin{equation}
T_q[\rho_n]:=\frac{1}{q-1}\left(1-W_q[\rho_n]\right),\; q>0,\; q\neq 1.
\label{eq:tsallis_n}
\end{equation}

The latter quantity is non-negative, extremal at equiprobability, concave for $q>0$ but pseudoadditive, while the Renyi entropies (also called by $q$-entropies in other contexts) are additive; moreover, $R_q[\rho_n]$ is a monotonically decreasing and pseudoconcave function of $q$ \cite{bector:nrlq86}. Both quantities reduce to the von Neumann or Shannon entropy for $q\to 1$.

From Eqs. (\ref{eq:wq_b}), (\ref{eq:Iq_th0}) and (\ref{eq:th0}), one finds that the entropic moments of the half-line Coulomb potential defined by Eq. (\ref{eq:wq_a}) have the expression
\begin{multline}
W_q[\rho_n]=\frac{Z^{q-1}}{2n^{3q-1}}\mathcal{I}_q\left[L_{n-1}^{(1)}\right]\\
=\frac{Z^{q-1}}{2n^{q-1}}\frac{(2q)!}{q^{2q+1}}
F_A^{(2q+1)}\left(2q+1,-n+1,\ldots,-n+1,0;2,\ldots,2,1;\frac{1}{q},\ldots,\frac{1}{q},1\right)\\
=\frac{Z^{q-1}}{2n^{q-1}}\frac{(2q)!}{q^{2q+1}}
\sum_{m_1=0}^{n-1}\ldots\sum_{m_{2q}=0}^{n-1} \frac{(2q+1)_{m_1+\ldots+m_{2q}}(-n+1)_{m_1}\cdots(-n+1)_{m_{2q}}}{(2)_{m_1}\cdots(2)_{m_{2q}}}
\frac{q^{-m_1}\cdots q^{-m_{2q}}}{m_1!\cdots m_{2q}!}.
\label{eq:wq_n}
\end{multline}

In particular, for $n=1$ and $2$ we have the following values
\begin{equation}
W_q[\rho_1]=\frac{Z^{q-1}(2q)!}{2q^{2q+1}},
\label{eq:wq_n1}
\end{equation}
and
\[
W_q[\rho_2]=\frac{Z^{q-1}}{2^{q}}\frac{(2q)!}{q^{2q+1}}\sqrt{\frac{2q}{\pi}}e^{-q} i K_{2q+\frac12}(-q),
\]
for the entropic moments of the ground and first excited state of our system, respectively. Let us point out that the ground-state value (\ref{eq:wq_n1}) has been recently obtained \cite{omiste:jmp09}.

Then, from Eqs. (\ref{eq:renyi_n}) and (\ref{eq:wq_n}) one readily has the expression
\begin{multline*}
R_q[\rho_n]=\frac{1}{1-q}\ln \left(\frac{Z^{q-1}}{2n^{q-1}}\frac{(2q)!}{q^{2q+1}}\right.\\
\left.\times
F_A^{(2q+1)}\left(2q+1,-n+1,\ldots,-n+1,0;2,\ldots,2,1;\frac{1}{q},\ldots,\frac{1}{q},1\right)\right)\\
=\frac{1}{1-q}\ln\left(\frac{Z^{q-1}}{2n^{q-1}}\frac{(2q)!}{q^{2q+1}}\right.\\
\left.\times
\sum_{m_1=0}^{n-1}\ldots\sum_{m_{2q}=0}^{n-1} \frac{(2q+1)_{m_1+\ldots+m_{2q}}(-n+1)_{m_1}\cdots(-n+1)_{m_{2q}}}{(2)_{m_1}\cdots(2)_{m_{2q}}}
\frac{q^{-m_1}\cdots q^{-m_{2q}}}{m_1!\cdots m_{2q}!}
\right),
\end{multline*}
for the $q$th-order Renyi entropy of the physical state of the half-line Coulomb potential characterized by the quantum number $n$. In a similar way, from Eqs. (\ref{eq:tsallis_n}) and (\ref{eq:wq_n}) we obtain
\begin{multline*}
T_q[\rho_n]=\frac{1}{q-1}\left(1-\frac{Z^{q-1}}{2n^{q-1}}\frac{(2q)!}{q^{2q+1}}\right.\\
\left.\times F_A^{(2q+1)}\left(2q+1,-n+1,\ldots,-n+1,0;2,\ldots,2,1;\frac{1}{q},\ldots,\frac{1}{q},1\right)\right)\\
=\frac{1}{q-1}\left(1-\frac{Z^{q-1}}{2n^{q-1}}\frac{(2q)!}{q^{2q+1}}\right.\\
\left.\times
\sum_{m_1=0}^{n-1}\ldots\sum_{m_{2q}=0}^{n-1} \frac{(2q+1)_{m_1+\ldots+m_{2q}}(-n+1)_{m_1}\cdots(-n+1)_{m_{2q}}}{(2)_{m_1}\cdots(2)_{m_{2q}}}
\frac{q^{-m_1}\cdots q^{-m_{2q}}}{m_1!\cdots m_{2q}!}
\right),
\end{multline*}
for the $q$th-order Tsallis entropy.

For the particular cases $n=1$ and $n=2$, we obtain the values
\[
R_q[\rho_1]=\frac{1}{1-q}\ln\frac{Z^{q-1}(2q)!}{2q^{2q+1}}
\]
\[
T_q[\rho_1]=\frac{1}{q-1}\left(1-\frac{Z^{q-1}(2q)!}{2q^{2q+1}}\right)
\]
for the $q$th-order ground-state Renyi and Tsallis entropies, respectively, and
\[
R_q[\rho_2]=\frac{1}{1-q}\ln\left(\frac{Z^{q-1}}{2^{q}}\frac{(2q)!}{q^{2q+1}}\sqrt{\frac{2q}{\pi}}e^{-q} i K_{2q+\frac12}(-q)\right)
\]
\[
T_q[\rho_2]=\frac{1}{1-q}\left(1-\frac{Z^{q-1}}{2^{q}}\frac{(2q)!}{q^{2q+1}}\sqrt{\frac{2q}{\pi}}e^{-q} i K_{2q+\frac12}(-q)\right)
\]
for the  $q$th-order Renyi and Tsallis entropies of the first excited state, also respectively.

\section{Disequilibrium and linear entropy of the half-line Coulomb potential}
\label{sec4}

This section is devoted to an alternative method of computation of the second-order entropic moment of the nth-state of the half-line Coulomb potential $W_2[\rho_n]$, also called disequilibrium because it measures the distance from the equilibrium state of the system; it receives other names in different contexts, such as Onicescu information energy, Heller or collision measure and inverse participation ratio. Moreover, it is closely connected with the linear entropy $L[\rho_n]\equiv L_n$ of the system, since $L_n= 1-\langle\rho_n\rangle$. The latter notion plays a very important role in quantum information theory because it is not only a measure of ``impurity'' of a quantum state but also it has been recently used as a successful measure of decoherence (see e. g. \cite{mokarzel:pra02}), entanglement \cite{dodonov:pla02, naudts:pra07, buscemi:pra07}, complexity \cite{catalan:pre02, lopez:pla95, sugita:pre02,anteneodo:pla96} and mixedness (see e. g. \cite{ghosh:pra01})  of quantum systems. Although this quantity can be calculated by putting $q=2$ in Eq. (\ref{eq:wq_n}), here we describe a method which give a much simpler and more operational expression.

According to Eq. (\ref{eq:wq_a}), the disequilibrium is given by:
\begin{equation}
\langle\rho_n\rangle=\frac{Z}{2n^{5}}\int_0^\infty t^{4}e^{-2t} \left(L_{n-1}^{(1)}(t)\right)^{4}dt.
\label{eq:evrhon}
\end{equation}

To calculate this integral we can use the general expression (\ref{eq:Iqgeneral}) or, alternatively, the method which is described in the following. First, we use the expansion
\[
\left(L_{n-1}^{(1)}(t)\right)^2=\frac{n}{2^{2(n-1)}}\sum_{k=0}^{n-1}\binom{2n-2-2k}{n-1-k}\frac{(2k)!}{k!(k+1)!}L_{2k}^{(2)}(2t),
\]
whose square is
\begin{multline}
\left(L_{n-1}^{(1)}(t)\right)^4=\frac{n^2}{2^{4(n-1)}}\left[\sum_{k=0}^{n-1}\binom{2n-2-2k}{n-1-k}^2
\left(\frac{(2k)!}{k!(k+1)!}\right)^2 \left(L_{2k}^{(2)}(2t)\right)^2\right.\\
\left.+2\sum_{k>k'}^{n-1}\binom{2n-2k-2}{n-k-1}\binom{2n-2k'-2}{n-k'-1}\frac{(2k)!}{k!(k+1)!} \frac{(2k')!}{k'!(k'+1)!}L_{2k}^{(2)}(2t)L_{2k'}^{(2)}(2t)\right].
\label{eq:L4}
\end{multline}

The combined use of Eqs. (\ref{eq:evrhon}) and (\ref{eq:L4}) involves the two following Laguerre functionals,
\[
\int_0^\infty t^4 e^{-2t}\left(L_{2k}^{(2)}(2t)\right)^2dt=\frac{1}{2^5}\int_0^\infty t^2 e^{-t} \left(tL_{2k}^{(2)}(t)\right)^2dt.
\]
and
\[
\int_0^\infty t^4 e^{-2t}L_{2k}^{(2)}(2t)L_{2k'}^{(2)}(2t)dt=\frac{1}{2^5}\int_0^\infty t^2 e^{-t} tL_{2k}^{(2)}(t) tL_{2k'}^{(2)}(t)dt.
\]
which may be evaluated by making use of the three-term recurrence relation \cite{temme_96}
\[
tL_{2k}^{(2)}(t)=(4k+3)L_{2k}^{(2)}(t)-(2k+1)L_{2k+1}^{(2)}(t)-(2k+2)L_{2k-1}^{(2)}(t).
\]
We obtain that
\[
\int_0^\infty t^4 e^{-2t}\left(L_{2k}^{(2)}(2t)\right)^2dt=\frac{(2k+1)(2k+2)}{2^5}\left[(4k+3)^2+(2k+1)(2k+3)+2k(2k+2)\right],
\]
and
\begin{multline*}
\int_0^\infty t^4 e^{-2t}L_{2k}^{(2)}(2t)L_{2k'}^{(2)}(2t)dt=\frac{1}{2^5}\left[\delta_{k,k'}\int_0^\infty t^2 e^{-t} \left(tL_{2k}^{(2)}(t)\right)^2dt\right.\\
+\delta_{k,k'-1}(2k+1)(2k+2)(2k+3)(2k+4)+\delta_{k,k'+1}(2k-1)2k(2k+1)(2k+2)\bigg]\\
=\frac{1}{2^5}(2k-1)2k(2k+1)(2k+2)\delta_{k,k'+1},\quad \text{for }k>k'.
\end{multline*}

Then, simple algebraic operations lead to
\begin{multline*}
\int_0^\infty t^4 e^{-2t}\left(L_{n-1}^{(1)}(t)\right)^4dt=\frac{n^2}{2^{4n-2}}\\
\times\sum_{k=0}^{n-1}\binom{2n-2k-2}{n-k-1}^2\left(\frac{(2k)!}{k!(k+1)!}\right)^2 \frac{(k+1)^2(2k+1)(3n-4k(2k-2n+1))}{n-k}.
\end{multline*}

Finally, the substitution of this expression into Eq. (\ref{eq:evrhon}) gives rise to the following value for the disequilibrium of the physical states of the half-line Coulomb potential
\begin{equation}
\langle\rho_n\rangle=Z D(n),
\label{eq:ev_rho}
\end{equation}
where
\begin{equation}
D(n)=\frac{1}{2^{4n-1}n^3} \sum_{k=0}^{n-1}\binom{2n-2k-2}{n-k-1}^2 \binom{2k}{k}^2 \frac{(2k+1)(3n-4k(2k-2n+1))}{n-k}.
\label{eqn:d_n}
\end{equation}

So, for example, the disequilibrium has the values
\[
\langle \rho_1\rangle=\frac{3Z}{8},\quad \langle\rho_2\rangle =\frac{33Z}{256},\quad \text{and } \langle\rho_3\rangle=\frac{17Z}{256},
\]
for the three lowest states of the system. It is worthwhile remarking that Eqs. (\ref{eq:ev_rho})-(\ref{eqn:d_n}) give a much simpler and more useful expression for the disequilibrium (and consequently for the linear entropy) of our system than Eq. (\ref{eq:wq_n}) with $q=2$.

\section{Shape complexity}
\label{sec5}

Here we discuss the LMC or shape complexity of the physical states of the half-line Coulomb potential, with emphasis on the lowest-lying and highest-lying (Rydberg) states of the system. This quantity \cite{catalan:pre02, lopez:pla95} is equal to the disequilibrium of the system times the Shannon entropic power; that is
\begin{equation}
C[\rho_n]=\langle\rho_n\rangle e^{S[\rho_n]},
\label{eq:shapecomplexity}
\end{equation}
where
\[
S[\rho_n]=-\int_0^\infty \rho_n(x)\ln\rho_n(x) dx
\]
denotes the renowned Shannon information entropy. The shape complexity occupies a very special position among the composite information-theoretic measures in general, as well as among the complexity measures. This is because of its three following properties: (a) invariance under replication, translation and rescaling transformations, (b) minimal value for the simplest probability densities (uniform and Dirac's delta), and (c) simple mathematical structure.

To calculate the shape complexity for a general physical state of our system we need to know not only its disequilibrium $\langle\rho_n\rangle$ (whose exact value has been obtained in the previous section) but also to find its Shannon entropy $S_n$. The latter quantity has been recently expressed \cite{omiste:jmp09} as
\begin{equation}
S[\rho_n]=3n+3\ln n-2\psi(n)-\frac{1}{n}-\frac{1}{2n}E_1\left[L_{n-1}^{(1)}\right]-2-\ln Z,
\label{eq:Srhon}
\end{equation}
where $\psi(n)$ is the psi or digamma function \cite{temme_96} and $E_1\left[L_{n-1}^{(1)}\right]$ is the Laguerre logarithmic functional
\begin{equation}
E_1\left[L_{n-1}^{(1)}\right]=\int_0^\infty t\omega_1(t) \left[L_{n-1}^{(1)}(t)\right]^2 \ln \left[L_{n-1}^{(1)}(t)\right]^2 dt,
\label{eqn:loglag}
\end{equation}
whose exact value has not yet been found; in fact, its calculation is a formidable task, not yet analytically accomplished for any generic quantum number $n$ except for its lowest and highest values. For the ground state ($n=1$) this logarithmic integral vanishes, so that
\[
S[\rho_1]=2\gamma-\ln Z \approx 1.1544 -\ln Z,
\]
where $\gamma\simeq 0.577$ is the Euler-Mascheroni constant.
For the high-lying or Rydberg ($n\gg 1$) states, the logarithmic integral (\ref{eqn:loglag}) has the asymptotic value \cite{dehesa:jmp98}
\[
E_1\left[L_{n-1}^{(1)}\right]=2n^2(3n-\ln n-\ln (2\pi)+o(1)),
\]
so that the corresponding Shannon entropy is, according to Eq. (\ref{eq:Srhon}),
\[
S[\rho_{Ry}]\equiv S[\rho_{n\gg 1}]=\ln\left(\frac{2\pi n^2}{Z e^2}\right)+o(1).
\]
Then, Eqs. (\ref{eq:ev_rho}), (\ref{eqn:d_n}), (\ref{eq:shapecomplexity}) and (\ref{eq:Srhon}) allows us to express the shape complexity of a generic quantum state $n$ of the half-line Coulomb potential as
\begin{equation}
C[\rho_n]=n^3D(n)\exp(A_n),
\label{eq:Crhon}
\end{equation}
where $D(n)$ is given by (\ref{eqn:d_n}) and $A_n$ is
\[
A(n)=3n-2\psi(n)-\frac{1}{n}-\frac{1}{2n}E_1\left(L_{n-1}^{(1)}\right)-2.
\]
It is worthwhile noting that the complexity does not depend on the potential strength $Z$. This, in fact, is true not only for the half-line Coulomb potential but also for any homogeneous potential in the sense recently pointed out by S.H. Patil et al \cite{patil:jpb07}. From Eq. (\ref{eq:Crhon}) we have, for instance, the value
\[
C[\rho_1]=\frac{3}{8}e^{2\gamma}\approx 1.1896,
\]
for the ground state. As well, we obtain the value
\[
C[\rho_{\text{Ry}}]\equiv C[\rho_{n\gg1}] \approx\frac{2\pi n^2}{e^2}D(n)
\]
for the shape complexity of the Rydberg states.

\section{Sharp upper bounds to Shannon entropy and shape complexity}
\label{sec6}

The explicit expressions of the Shannon entropy $S[\rho_n]$ and the shape complexity $C[\rho_n]$ were not given in the previous Section because of the mathematical difficulties to evaluate the Laguerre functional $E_1(L_{n-1}^{(1)})$ analytically in Eqs. (\ref{eq:Srhon})-(\ref{eqn:loglag}), except for the states at the two extremal regions of the energetic spectrum of our system. Then, it is natural to try to obtain sharp bounds to these two information-theoretic quantities for all physical states of the half-line Coulomb potential. Here, we do this task by taking into account the non-negativity of the Kullback-Leibler entropy of the two arbitrary probability densities $\rho(x)$ and $f(x)$,
\[
I_{\rm KL}[\rho,f]=\int_0^\infty \rho(x)\ln\frac{\rho(x)}{f(x)}dx,
\]
followed by an optimization procedure. The former step applied to the quantum-mechanical density $\rho_n(x)$ of our system and the prior density $f(x)$, duly normalized to unity, leads to the upper bound
\[
S[\rho_n]\leq -\int_0^\infty \rho_n(x)\ln f(x)dx,
\]
In particular, with the following choice for the prior density
\[
f(x)=\frac{ka^{\frac{1}{k}}}{\Gamma\left(\frac{1}{k}\right)}e^{-ax^k},\; k=1,2,\ldots
\]
and optimizing the result with respect to $a$, one has that
\[
S[\rho_n]\leq \ln\left[A_k\left\langle x^k\right\rangle_n^{\frac{1}{k}}\right],\; k=1,2,\ldots
\]
with the optimal value $a_{\rm opt}=(k\langle x^k\rangle_n)^{-1}$ and the constant
\[
A_k=\frac{(ek)^{\frac{1}{k}}}{k}\Gamma\left(\frac{1}{k}\right).
\]
The $k$th-order expectation value of the quantum-mechanical probability density $\rho_n(x)$  given by (\ref{eq:rho}) is \cite{omiste:jmp09}
\begin{equation}
\left\langle x^k\right\rangle_n:=\int_0^\infty x^k\rho_n(x)dx=\frac{B_k(n)}{Z^k}
\label{eq:ev_xk}
\end{equation}
with

\begin{equation}
B_k(n)=\frac{n^{k-2}}{2^{k+1}}\sum_{i=0}^{n-1}\binom{k+1}{n-i-1}^2\frac{(k+i+2)!}{i!};\; k=1,2,\ldots
\label{eqn:bk_n}
\end{equation}

Then, e.g. for $k=1$ and $2$ one has that
\[
\langle x\rangle_n=\frac{3n^2}{2Z}\quad\text{and}\quad \langle x^2\rangle_n=\frac{n^2(5n^2+1)}{2Z^2}.
\]
Consequently, the Shannon entropy of the half-line Coulomb potential is bounded as
\begin{equation}
S[\rho_n]\leq \ln\left\{\frac{A_k}{Z}\left[B_k(n)\right]^\frac{1}{k}\right\}\equiv b(k,n); \;   k=1, 2, ...
\label{eq:shannon_bound}
\end{equation}
Taking into account the explicit values (\ref{eq:ev_rho}) and (\ref{eq:ev_xk}) for the quantities $\langle \rho_n\rangle$ and $\langle x^k\rangle_n$, we can readily obtain the inequality
\begin{equation}
C[\rho_n]\le A_k D(n)\left[B_k(n)\right]^\frac{1}{k}\equiv c(k,n);\; k=1,2,\ldots
\label{eq:c_bound}
\end{equation}
which provides a set of upper bounds to the shape complexity of the physical state characterized by the quantum number $n$ and the order $k$ of the expectation value. $D(n)$ and $B_k(n)$ are given by Eqs. (\ref{eqn:d_n}) and (\ref{eqn:bk_n}), respectively.

\begin{figure}
\begin{center}
\includegraphics[width=10cm]{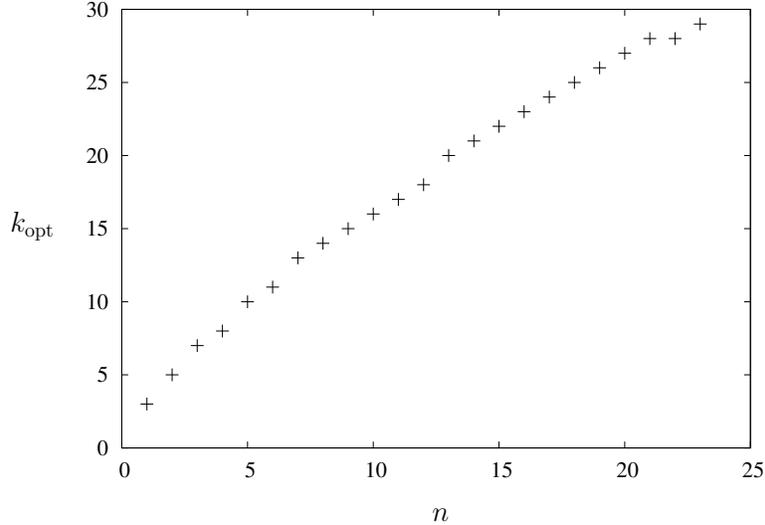}
\end{center}
\caption{Values of $k=k_{\rm opt}$ giving the best upper bounds $b(k,n)$ to the Shannon entropy and $c(k,n)$ to the complexity for a few low-lying quantum states of the half-line Coulomb potential.}
\label{fig:kopt}
\end{figure}

\begin{figure}
\begin{center}
\includegraphics[width=10cm]{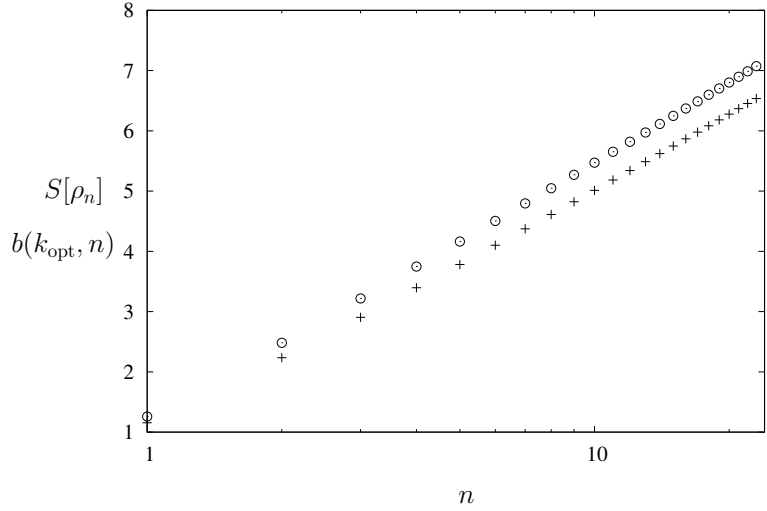}
\end{center}
\caption{Study of the accuracy of the upper bound $b(k_{\rm opt},n)$ ($\circ$) to the Shannon entropy $S[\rho_n]$ ($+$) for various low-lying states of the half-line Coulomb potential}
\label{fig:shabound}
\end{figure}

\begin{figure}
\begin{center}
\includegraphics[width=10cm]{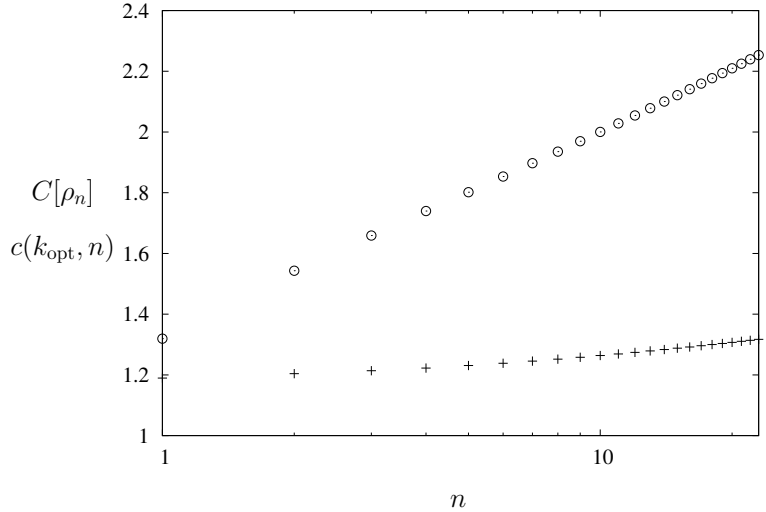}
\end{center}
\caption{Study of the accuracy of the upper bound $c(k_{\rm opt},n)$ ($\circ$) to the complexity $C[\rho_n]$ ($+$) for various low-lying states of the half-line Coulomb potential.}
\label{fig:combound}
\end{figure}

Finally, we make a numerical study of the accuracy of the Shannon and complexity inequalities (\ref{eq:shannon_bound}) and (\ref{eq:c_bound}), respectively. First we calculate the values $k_{\rm opt}$ which provide the best bound to the Shannon entropy and complexity of our system. They are given in Figure \ref{fig:kopt}.  Then, the accuracy of the best bounds $b(k_{\rm opt},n)$ to the Shannon entropy and $c(k_{\rm opt},n)$ to the complexity for the first few low-lying quantum states of the half-line Coulomb potential is studied in Figures \ref{fig:shabound} and \ref{fig:combound} by means of their comparison with the corresponding numerical values, respectively. We observe that $k_{\rm opt}$ grows when the system is more and more excited. This is because the prior density $f(x)$ needs a larger value of $k$, as $n$ grows, to accommodate itself to the extension of $\rho_n(x)$, making the Kullback-Leibler entropy minimal. Thus, the best upper bound is given by an increasing value of $k$ as the system gets more and more excited. Moreover, the bounds $b(k_{\rm opt},n)$ and $c(k_{\rm opt},n)$ to the Shannon entropy and complexity have a monotonic behaviour similar to the exact values of these two quantities, respectively, but the corresponding relative errors are opposite. Indeed, while the relative error of the Shannon bound decreases very fast when $n$ is increasing, the opposite occurs for the complexity bound.

\section{Information-theoretic lengths}
\label{sec7}

There exist \cite{hall:pra99,hall:pra00} different information-theoretic measures of spreading which share some relevant properties with the familiar root-mean-square or standard deviation $(\Delta x)_n=\sqrt{\langle x^2\rangle_n-\langle x\rangle_n^2}$; namely, invariance under translations and reflections, linear scaling in the variable (so that $\Delta y=\Delta x/\lambda$ for $y=\lambda x$) and dimensions of length. These spreading measures are defined as

\begin{equation}
L_q^R[\rho_n]=\exp\left(R_q[\rho_n]\right)=\langle [\rho_n(x)]^{q-1}\rangle^{-\frac{1}{q-1}}=\left[W_{q}[\rho_n]\right]^{-\frac{1}{q-1}}
\label{eqn:lqr}
\end{equation}
for the Renyi length of order $q$,
\[
L^S[\rho_n]=\exp\left(S[\rho_n]\right)
\]
for the Shannon length, and
\[
(\delta x)_n=\frac{1}{\sqrt{F[\rho_n]}}, \quad \text{where}\quad F[\rho_n]=\int_0^\infty dx \frac{[\rho'_n(x)]^2}{\rho_n(x)},
\]
for the Fisher length. They quantify the spreading of the quantum-mechanical probability cloud in a complementary and qualitatively different way. The Renyi and Shannon lengths have a global character as the standard deviation because they are powerlike (Renyi, standard deviation) and logarithmic (Shannon) functionals of the probability density, respectively. Moreover, they provide quantifiers of the extent to which the distribution is in fact concentrated, rather than a measure of the separation of the region of concentration from the centroid or mean value, a particular point of the distribution. In contrast, the Fisher length is a local measure of spreading because it is a gradient functional of the density; it measures the concentration of the distribution between its nodes, so quantifying the oscillating or wiggliness character of the corresponding wavefunction.

Recently, we have found \cite{omiste:jmp09} the values

\[
(\Delta x)_n=\frac{n}{Z}\sqrt{n^2+2}, \hspace{1cm} (\delta x)_n=\frac{n}{2Z}; \hspace{.5cm} n=1, 2, ...
\]
for the standard deviation and the Fisher length, respectively, of all quantum states of the half-line Coulomb potential.
Moreover, we were able to calculate \cite{omiste:jmp09} the Renyi length for the ground state of the system obtaining
\begin{equation}
L_q^R[\rho_1]=\frac{q^{2+\frac{3}{q-1}}}{Z}\left(\frac{2}{\Gamma(2q+1)}\right)^{\frac{1}{q-1}},
\label{eqn:lqr1}
\end{equation}
and the Shannon length for the extremal states of the energetic spectrum of our system; namely, the ground state $\rho_1(x)$ and the Rydberg states $\rho_{Ry}(x)\equiv\rho_{n\gg1}(x)$ in which case we have the values 
\[
L^S[\rho_1]=\frac{e^{2\gamma}}{Z},\qquad L^S[\rho_{Ry}]\simeq\frac{2\pi n^2}{Ze^2},
\]
where $\gamma\simeq 0.577$ is the Euler-Mascheroni constant.

\begin{figure}
\begin{center}
\includegraphics[width=10cm]{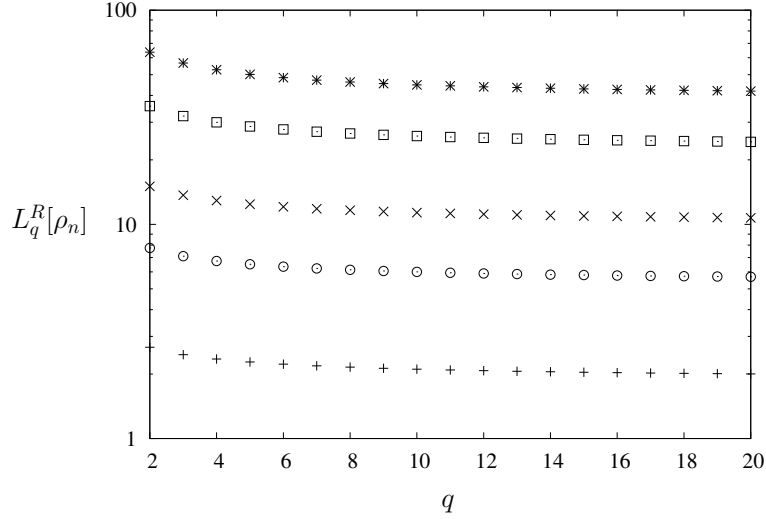}
\end{center}
\caption{Renyi lengths $L_q^R[\rho_n]$ in terms of the order $q$ for various quantum states of the half-line Coulomb potential with $n=1$ ($+$), 2 ($\circ$), 3 ($\times$), 5 ($\square$), 7 ($*$), and $Z=1$}
\label{fig:renyi_length_q}
\end{figure}

\begin{figure}
\begin{center}
\includegraphics[width=10cm]{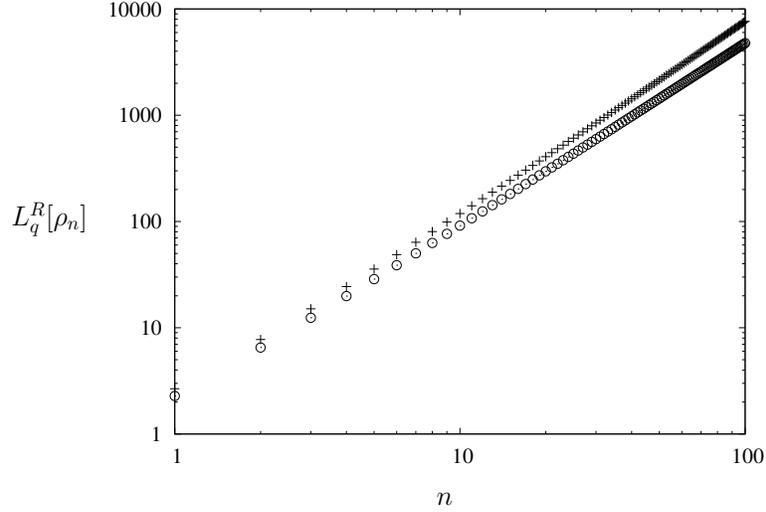}
\end{center}
\caption{Renyi lengths $L_q^R[\rho_n]$ with $q=2$ ($+$), and $q=5$ ($\circ$), for the first 100 states of the half-line Coulomb potential with $Z=1$}
\label{fig:renyi_length_n}
\end{figure}

Here we find the Renyi lengths $L_q^R[\rho_n]$ for an arbitrary state of the half-line Coulomb potential characterized by the probability density $\rho_n(x)$. Moreover, we study numerically both uncertainty measures in terms of $q$ (Renyi length) and $n$ (Renyi and Shannon lengths). Let us start noting that Eqs. (\ref{eq:wq_a}), (\ref{eq:wq_b}) and (\ref{eqn:lqr}) yield the value
\[
L_q^R[\rho_n]=\frac{n^\frac{3q-1}{q-1}}{Z}\left[\frac{2}{\mathcal{I}_q\left[L_{n-1}^{(1)}\right]}\right]^{\frac{1}{q-1}}
\]
for the Renyi length, where $\mathcal{I}_q\left[L_{n-1}^{(1)}\right]$ is given by Eq. (\ref{eq:Iqgeneral}). In particular, for $n=1$ and $n=2$ we obtain the expression (\ref{eqn:lqr1}) for the Renyi length $L_q^R[\rho_1]$ of the ground state and the value
\[
L_q^R[\rho_2]=\frac{2^\frac{q}{q-1}}{Z}\left[\frac{q^{2q+1}}{(2q)! \sum_{j=0}^{2q}\binom{2q}{j}\frac{(2q+1)_j (-1)^j}{2^j q^j}}\right]^{\frac{1}{q-1}}
\]
for the Renyi length of the first excited state. In calculating these values we have used the expressions (\ref{eqn:iq0}) and (\ref{eqn:i1}) for the entropic functionals of the involved Laguerre polynomials. The Renyi lengths $L_q^R[\rho_n]$ are numerically studied in Figures \ref{fig:renyi_length_q} and \ref{fig:renyi_length_n} as functions of the order $q$ (for five excited states) and the quantum number $n$ (for two fixed values of $q$), respectively. Therein, we have assumed without any loss of generality that $Z=1$. We observe that
\begin{eqnarray*}
L_{q_1}^R[\rho_n]&<&L_{q_2}^R[\rho_n], \quad \text{for } q_1>q_2,\\
L_q^R[\rho_{n_1}]&<&L_q^R[\rho_{n_2}], \quad \text{for }  n_1<n_2.
\end{eqnarray*}
As expected, the Renyi lengths increase with $n$, since the wavefunction expands. However, as a function of $q$, they present a monotonic decreasing behaviour, tending to a constant value for large values of $q$ what seems to indicate a saturation of the information given by this measure.

\begin{figure}
\begin{center}
\includegraphics[width=10cm]{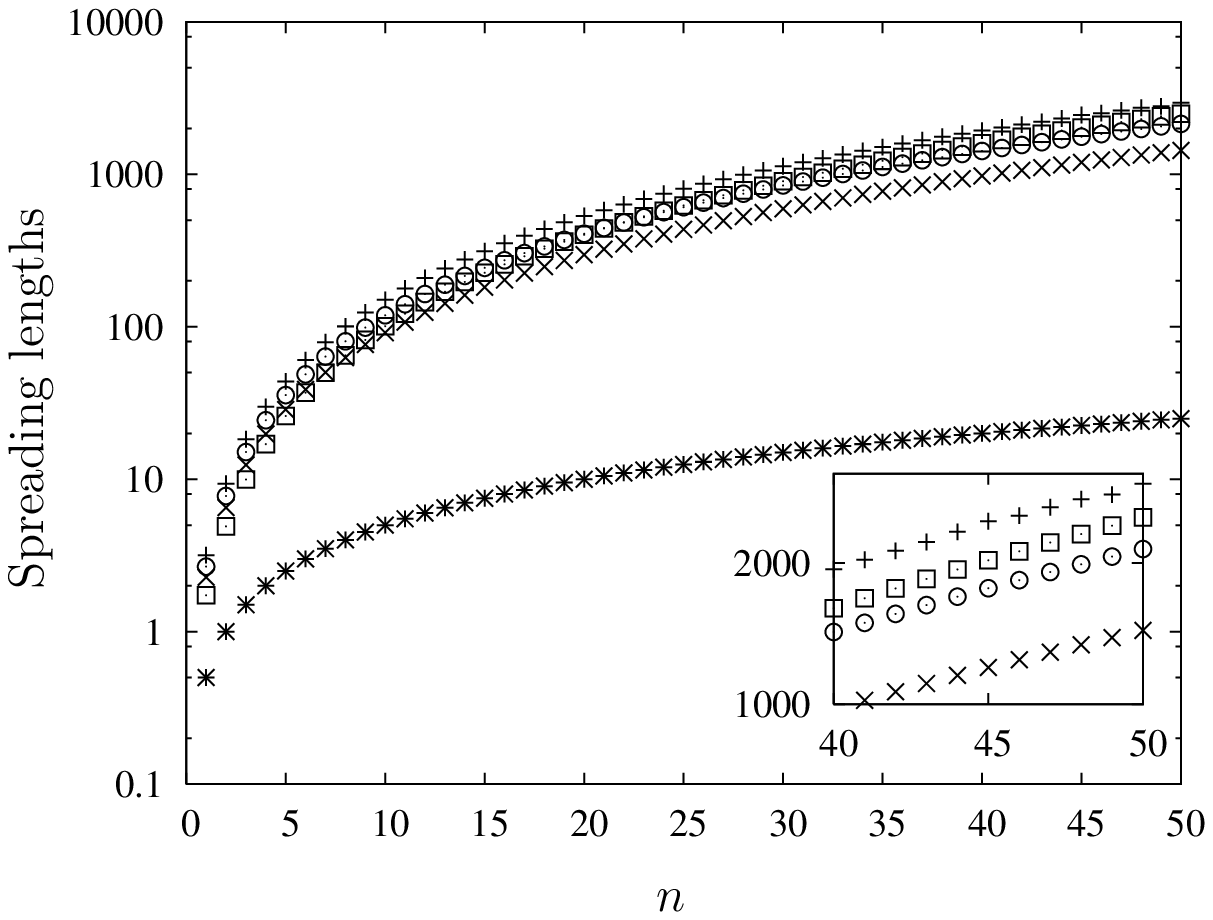}
\end{center}
\caption{Comparison of the Shannon length ($+$), standard deviation ($\square$), Renyi length for $q=2$ ($\circ$), Renyi length for $q=5$ ($\times$) and Fisher length ($*$) for the first 50 states of the half-line Coulomb potential with $Z=1$. A detailed view for the states with $n$ from 40 to 50 is given in the inset.}
\label{fig:lengths}
\end{figure}

Moreover, in Figure \ref{fig:lengths} we make the comparison of the Renyi and Shannon lengths with the standard deviation and the Fisher length, as given by Eq. (\ref{eqn:lqr1}) for the first 50 quantum states of our system. We observe that for states with $n\gtrsim 20$, the following relations are satisfied
\[
(\delta x)_n < L_q^R[\rho_n] < (\Delta x)_n < L^S[\rho_n]
\]
which manifest that the Fisher length is the proper measure of uncertainty for 1D-hydrogenic systems.

\section{Conclusions}

First, the linear Renyi and Tsallis entropies of the half-line Coulomb system have been analytically calculated. These information-theoretic quantities are closely connected with the entropic moments of the system, which essentially depend on some entropic functionals of the Laguerre polynomials that control the wavefunctions of its stationary states. These functionals have been determined by the non-sufficiently known linearization theorem of Srivastava-Niukkanen.

Then, the Shannon entropy and the shape complexity of this system are explicitly given for the ground state and the very high-lying (i.e., Rydberg) states. Moreover, sharp bounds to these two quantities are found for arbitrary states of the system and their accuracy is discussed in detail. Most interesting is the mutual comparison among all the known direct measures of spreading of the charge distribution of the system; namely, the standard deviation and the information-theoretic lengths of Renyi, Shannon and Fisher lengths.

Finally, the uncertainty measures of the half-line Coulomb system defined by its information-theoretic lengths, as well as their associated entropic moments, are numerically discussed and compared with the previously known standard deviation and Fisher length \cite{omiste:jmp09}

\section{Acknowledgements}

This work has been partially supported by the Spanish MICINN grant FIS2008-02380, and the grants FQM-1735 and FQM-2445 of Junta de Andaluc\'{\i}a. The authors belong to the research group FQM-207. J.J. Omiste acknowledges the scholarship ``Iniciaci\'on a la investigaci\'on'' in the framework of the Plan Propio of the University of Granada.

\providecommand{\bysame}{\leavevmode\hbox to3em{\hrulefill}\thinspace}
\providecommand{\MR}{\relax\ifhmode\unskip\space\fi MR }
\providecommand{\MRhref}[2]{%
  \href{http://www.ams.org/mathscinet-getitem?mr=#1}{#2}
}
\providecommand{\href}[2]{#2}

\end{document}